# Ultrafast manipulation of the NiO antiferromagnetic order via sub-gap optical excitation

Xiaocui Wang,[a] Robin Y. Engel,[bc] Igor Vaskivskyi,[d] Diego Turenne,[a]
Vishal Shokeen,[a] Alexander Yaroslavtsev,[a] Oscar Grånäs,[a] Ronny Knut,[a]
Jan O. Schunck,[bc] Siarhei Dziarzhytski,[b] Günter Brenner,[b] Ru-Pan Wang,[c]
Marion Kuhlmann,[b] Frederik Kuschewski,[b] Wibke Bronsch,[e]
Christian Schüßler-Langeheine,[f] Andriy Styervoyedov,[g] Stuart S. P. Parkin,[g]
Fulvio Parmigiani,[e] Olle Eriksson,[a] Martin Beye[bc] and Hermann A. Dürr*[a]

Wide-band-gap insulators such as NiO offer the exciting prospect of coherently manipulating electronic correlations with strong optical fields. Contrary to metals where rapid dephasing of optical excitation via electronic processes occurs, the sub-gap excitation in charge-transfer insulators has been shown to couple to low-energy bosonic excitations. However, it is currently unknown if the bosonic dressing field is composed of phonons or magnons. Here we use the prototypical charge-transfer insulator NiO to demonstrate that 1.5 eV sub-gap optical excitation leads to a renormalised NiO band-gap in combination with a significant reduction of the antiferromagnetic order. We employ element-specific X-ray reflectivity at the FLASH free-electron laser to demonstrate the reduction of the upper band-edge at the O 1s-2p core-valence resonance (K-edge) whereas the antiferromagnetic order is probed via X-ray magnetic linear dichroism (XMLD) at the Ni 2p-3d resonance ($L_2$-edge). Comparing the transient XMLD spectral line shape to ground-state measurements allows us to extract a spin temperature rise of 65±5 K for time delays longer than 400 fs while at earlier times a non-equilibrium spin state is formed. We identify transient mid-gap states being formed during the first 200 fs accompanied by a band-gap reduction lasting at least up to the maximum measured time delay of 2.4 ps. Electronic structure calculations indicate that magnon excitations significantly contribute to the reduction of the NiO band gap.

[a] *Department of Physics and Astronomy, Uppsala University, 75120 Uppsala, Sweden*
[b] *Deutsches Elektronen Synchrotron DESY, Notkestr. 85, 22607 Hamburg, Germany*
[c] *Department of Physics, Universität Hamburg, Luruper Chaussee 149, 22761 Hamburg, Germany*
[d] *Department for complex matter, Jozef Stefan Institute, Jamova cesta 39, 1000 Ljubljana, Slovenija*
[e] *Elettra-Sincrotrone Trieste S.C.p.A., 34149 Basovizza, Italy*
[f] *Helmholtz-Zentrum Berlin für Materialien und Energie, Albert-Einstein-Str. 15, 12489 Berlin, Germany*
[g] *Max-Planck Institut für Mikrostrukturphysik, Weinberg 2, 06120 Halle, Germany*
\* E-mail: hermann.durr@physics.uu.se

# 1 Introduction

Antiferromagnets (AFMs) have gained increasing interest and attention in both fundamental research and technological development. Since AFMs have zero net magnetization in their ground state, they are robust against external magnetic fields and have the promising potential to facilitate spintronic devices with ultrahigh speeds in the THz range.[1,2] Despite intense experimental studies and theoretical modelling efforts that have been dedicated to the coupling mechanisms between charge, spin and lattice degrees of freedom, the important interaction pathways remain elusive.

Owing to the development of ultrafast optical laser and free-electron laser (FEL) radiation sources that improve the temporal resolution even towards the few-femtosecond regime and the development of computing algorithms, scholars in this field have been able to unambiguously detect and model the complex dynamics in strongly correlated materials. With scattering techniques, using electrons,[3] X-ray photons[4] or neutrons,[5] one can investigate the ordering of the lattice or the AFM order. The diffraction signal of multiple-integer Bragg peaks is assigned to the structural re-arrangement, while the multiples of half-integer peaks are assigned to the magnetic ordering.[6] With spectroscopic techniques, one can study the electronic configurations with element specificity[7-9] and spin excitations in magnetic systems.[10]

Nickel oxide (NiO), as a prototype AFM, is a good candidate for studies in strongly correlated materials due to its well separated intra-gap states,[11] large spin density[5,12,13] and a high Néel temperature ($T_N \sim 523$ K).[7,14,15] Above $T_N$, NiO has a rock-salt structure (point group Fm-3m) with a lattice constant of 4.176 Å. Below $T_N$, the $Ni^{2+}$ spins are ferromagnetically aligned along <11-2> directions within the (111) plane, and adjacent (111) planes are coupled antiferromagnetically, which composes two sublattice AFM systems. Due to exchange striction, the AFM ordering of the spins induces a small contraction along the <111> direction, resulting in a rhombohedral distortion of the crystalline structure. This distortion induces a reduction of the crystallographic symmetry from point group Fm-3m of the cubic structure to point group C2/m of the rhombohedral structure. Since there are four energetically degenerate <111> directions, the distortion can occur along any of the four equivalent directions. This results in four types of twin-domains (T-domain). For each T-domain, there are three possible spin orientations (S-domain). In total, there can be 12 orientational domains in NiO crystals.

In various static measurements, the typical X-ray magnetic linear dichroic (XMLD) line shape, which is directly linked to the long-range ordering in AFMs, has been observed in NiO by varying the temperature,[7] the experimental geometry[7,16] or measuring XMLD from different AFM domains in microscopy.[17,18] There have also been a large number of time-resolved measurements carried out in NiO in order to disentangle the coupling between different degrees of freedom. With time-resolved optical measurements, oscillations with THz frequencies have been observed and assigned to magnon excitation.[14,19,20] In a recent experiment, researchers have investigated the lattice dynamics of NiO using femtosecond electron diffraction,[3] and the origin of the rhombohedral lattice distortion was assigned to the weakening of the AFM order. However, there is so far no direct evidence of a laser-induced change in the AFM order driven by femtosecond laser pulses in time-resolved magnetic diffraction measurements.[6]

In this article, we report the investigation of the ultrafast manipulation of the AFM order in NiO films via sub-gap optical excitation. Using time-resolved resonant X-ray reflectivity, we observe XMLD line shape changes at the nickel $L_2$-edge, i.e., for $2p_{1/2} - 3d$ transitions, and band gap renormalisation at the oxygen K-edge (1s – 2p transitions). By comparing with static XMLD measurements at different temperatures,[7] we determine the laser-induced temperature change of the spin system to be 65 ± 5 K. In order to explain the band gap renormalisation observed at the oxygen K-edge, we use first principles calculations to investigate the influence of magnetic excitations on the band structure with the density-functional theory + Hubbard U formalism.

## 2 Methods

### 2.1 Sample preparation and characterization

NiO(001) crystalline films were used as our samples. The NiO films were epitaxially grown with surface normal along [001] direction on a single crystal MgO (001) substrate. The MgO substrate was polished on both sides. A 2-nm thick MgO underlayer was first deposited by radio frequency magnetron sputtering in 3 mTorr argon at a temperature below 100°C. The NiO layer was then deposited at 700°C in a gas mixture of Ar(90%)/$O_2$(10%) at a pressure of 3 mTorr and was annealed in-situ for 15 min with the same temperature and gas mixture condition. The thickness of the NiO film was 40 nm.

The ground-state electronic and magnetic structure of NiO crystalline films was characterised by performing X-ray absorption spectroscopy (XAS) measurements in drain current mode at the PM3 beamline at the BESSY II electron storage ring operated by Helmholtz-Zentrum Berlin.[21] Static XAS spectra were acquired for various incidence angles at room temperature.

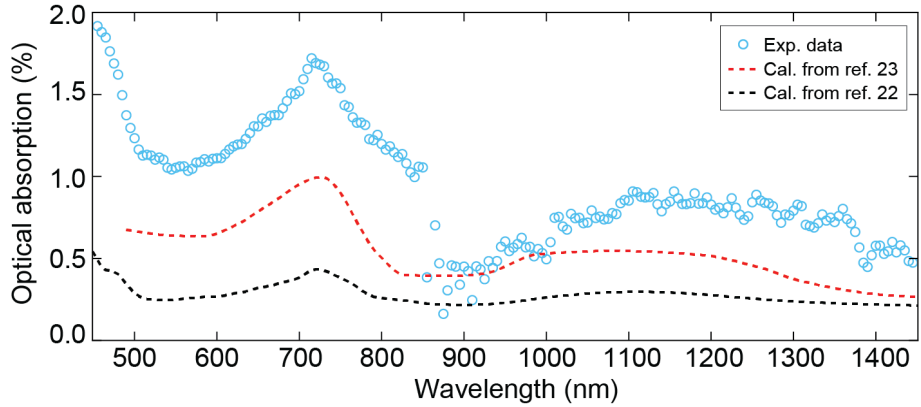

**Fig. 1** The optical absorption of the NiO film with a thickness of 40 nm at wavelengths between 450 nm and 1450 nm. The circles represent the experimental data obtained from visible-infrared spectroscopy measurement. The sharp drop at 860 nm is due to the change of photodetector in the measurement. The dashed black and red lines denote the calculated optical absorption using the reported absorption coefficient in ref. [22] and ref. [23], respectively.

The previous optical measurements of NiO,[22,23] show a variation in the optical absorption coefficient. Therefore, optical spectroscopy measurements were carried out prior to the time-resolved X-ray reflectivity experiment in order to evaluate the optical absorption properties of the NiO film during the optical pump – X-ray probe experiments, as shown in Fig. 1. The symbols in Fig. 1 represent the measured optical absorption between 450 nm and 1450 nm, and the dashed lines represent the calculated absorption for a 40-nm NiO film using the tabulated absorption coefficient.[22,23] The sharp drop at 860 nm in the experimental data is due to the change of photodetector in the optical measurements. From experimental data one can see two absorption bands that are centred at 720 nm and 1150 nm. The peak position of these two absorption bands is close to the values reported in previous works.[22-24] The origin of these two absorption bands has been assigned to the crystal-field d-d transitions.[25,26] We note that the dashed red line is truncated at 500 nm because the short wavelength limit reported in ref. [23] is 500 nm. Overall, the absorption bands shown in the experimental data indicate the good quality of our sample.

### 2.2 Time-resolved X-ray reflectivity measurements at grazing angles

The time-resolved X-ray reflectivity experiments were carried out using the MUSIX end-station at beamline FL24 at the free-electron laser, FLASH, in Hamburg.[27,28] Experiments were carried out in reflection geometry as illustrated in Fig. 2(a). The NiO film was excited by a femtosecond laser pulse with a central wavelength around 800 nm and a pulse duration of 50 fs. The temporal evolution of the

X-ray reflectivity following laser excitation was probed at both the oxygen K-edge and the nickel $L_{2,3}$-edges. The angle between the incident pump and the probe beams was 0.75°, i.e. nearly collinear.

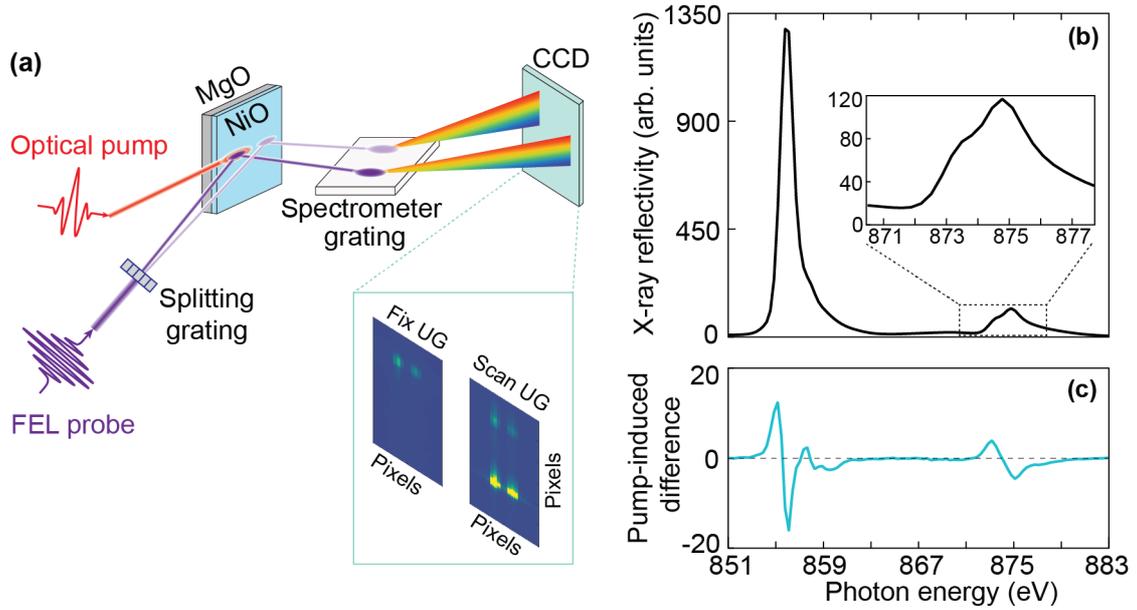

**Fig. 2** (a) Schematic of experimental set-up for time-resolved X-ray reflectivity measurements. The inset shows two representative CCD images for measurements with fixed undulator gap (UG) and scanning UG. (b) Static X-ray reflectivity spectrum of NiO at the nickel $L_{2,3}$-edges. The inset shows the enlarged spectrum at the nickel $L_2$-edge. (c) Pump-induced difference in the X-ray reflectivity spectrum for time delays longer than 400 fs.

FLASH was operated in a 10 Hz burst mode with each burst consisting of 40 X-ray pulses. The FEL was tuned for the third harmonic radiation to reach the oxygen K-edge and the nickel $L_{2,3}$-edges. Reflectivity spectra were recorded utilizing a split-beam normalisation scheme.[28] As illustrated in Fig. 2(a), the FEL probe beam propagated through a beam splitting grating and was horizontally split such that the zeroth and first diffraction orders hit the sample. The zeroth order beam overlapped both spatially and temporally with the optical pump beam to record the transient X-ray reflectivity. The zeroth and first diffraction orders were separated by 3 mm on the sample, ensuring that the first diffraction order of the FEL beam had no overlap with the optical laser beam and could be used to normalise fluctuations of the FEL spectral intensity due to the self-amplified spontaneous emission (SASE) process. X-ray energy resolution was obtained by directing both the zeroth and first diffraction orders onto a down-stream spectrometer grating and consecutively, a CCD camera. The dispersion of the spectrometer was calibrated and found to be 0.33 eV/pixel. To measure reflectivity spectra around the absorption edges, the photon energy was scanned in steps of 0.5 eV at the fundamental wavelength by varying the undulator gap, which corresponds to 1.5 eV at the third harmonic in the time-resolved X-ray reflectivity measurement. As each FEL spectrum covers a bandwidth of about 1% of the total photon energy, this ensures a seamless spectrum of the scanned range. In reflection geometry, by varying the angle of incidence, one can access different parts of the Brillouin zone. The size of the X-ray beam was 50 μm × 75 μm. The X-ray angle of incidence was set to 7.23° at the nickel $L_{2,3}$-edges and 9.00° at the oxygen K-edge in order to access a similar $q$-range in reciprocal space and match the X-ray penetration depth with the sample thickness. The X-ray penetration depth is 50 nm at the oxygen K-edge and 60 nm at the nickel $L_{2,3}$-edges.[29] The transferred wave-vector $q$ was 0.85 nm$^{-1}$ at oxygen K-edge and 1.09 nm$^{-1}$ at the nickel $L_{2,3}$-edges, which corresponds to 0.06 and 0.07 reciprocal lattice unit, respectively. Therefore, the $q$-range lies close to the Γ point of the Brillouin zone. Figure 2(b) shows a static X-ray reflectivity spectrum of NiO at the nickel $L_{2,3}$-edges, and the enlarged nickel $L_2$-edge is displayed in the inset.

The size of the optical pump beam was characterized both by imaging a virtual focus on an equivalent plane and by a knife-edge measurement. The size of the pump beam was (270 ± 10) μm × (257 ± 2) μm. The incident pump fluence was 60 mJ/cm$^2$ and 75 mJ/cm$^2$ due to the different angles of incidence at the nickel $L_{2,3}$-edges and the oxygen K-edge, respectively. Figure 2(c) shows the pump-induced change of the X-ray reflectivity spectrum at the nickel $L_3$- and $L_2$-edges averaged over time delays between 400 fs and 2.4 ps. We will describe in the following section 2.3 the data analysis procedure and involved uncertainties. The temporal resolution and pump-probe temporal overlap were characterized at regular intervals during the measurements by probing the transient optical reflectivity change of a GaAs single crystal induced by the FEL pulse.[30] The temporal resolution was found to be 300 ± 100 fs full-width half-maximum (FWHM), which mainly arises from the jitter between the pump and probe pulses. The temporal resolution evaluated from the GaAs single crystal is overestimated because the measurements at the oxygen K-edge reveal dynamics faster than 300 fs. Therefore, the GaAs single crystal was mainly used to monitor the temporal overlap, i.e., time zero. During the experiments, the delay between the optical laser and the FEL was varied by a delay stage in the optical path. The delay was scanned across a 3-ps time window in steps of 100 fs.

**2.3 Data analysis for the normalisation of the SASE FEL fluctuations and slow drifts**

In order to quantitatively extract the laser-induced reflectivity changes from spectra acquired at SASE FELs in an unambiguous manner, one needs to account for the shot-to-shot fluctuations in spectral intensity caused by the SASE process. Since the measured reflectivity spectrum is proportional to the multiplication of the FEL spectrum, the spectral reflectivity of the sample, and the efficiency of the diffraction grating, the direct measure of the laser-induced reflectivity change is the ratio of the measured spectra without and with laser excitation. To compensate for the slow drifts between the spectra acquired with and without laser excitation, spectral changes observed from the signal sample spot, which is illuminated by both the optical laser beam and the FEL beam, are normalised by the changes observed from the reference sample spot illuminated only by the FEL beam. The normalised ratio eventually gives the real pump-probe effect.

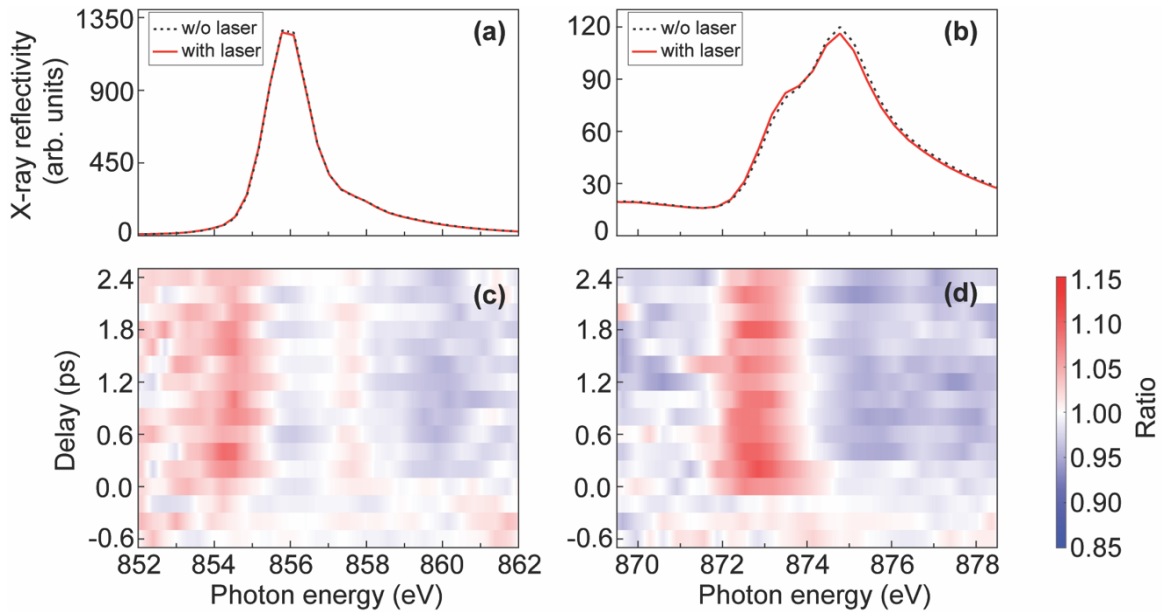

**Fig. 3** (a), (b) X-ray reflectivity spectra of NiO at the nickel $L_{2,3}$-edges. The dashed black and solid red lines represent the spectra acquired at time delays longer than 400 fs without and with laser excitation, respectively. (c), (d) The ratio map as a function of delay and photon energy at the nickel $L_{2,3}$-edges.

Figures 3(a, b) show the X-ray reflectivity spectra at the nickel $L_{2,3}$-edges, respectively. The dashed black and the solid red lines represent the spectra acquired without and with laser excitation, respectively. From the comparison in Figs. 3(a, b), one can directly see that the change at the nickel

$L_3$-edge is significantly smaller than at the nickel $L_2$-edge. To quantify the relative change induced by the optical laser, we calculated the ratio of the spectra with and without laser excitation. Figures 3 (c, d) show the ratio plots as a function of delay and photon energy at the nickel $L_{2,3}$-edges, respectively. In Figs. 3(c, d), it can be seen that the laser-induced change at the nickel $L_2$-edge is less noisy than that at the nickel $L_3$-edge. Therefore, in the results and discussion about the nickel L-edge measurements, we will focus mainly on the nickel $L_2$-edge.

## 3 Results and discussion

In this section we present time-resolved measurements at the O K-edge and Ni $L_{2,3}$-edges that probe O 1s–2p and Ni 2p-3d core-valence excitations, respectively. These orbitally resolved measurements enable us to extract information about the temporal evolution of the band-gap renormalisation and the modification of the AFM order, respectively. We will finally present a model that relates the changes of the upper band-gap, composed of O 2p and Ni 4s orbitals, to canted AFM moments that typically occur during magnon excitation. We start the section by describing the measured X-ray reflectivity spectra and relating them to XAS measurements via Kramers-Kronig analysis.

### 3.1 Comparison of X-ray reflectivity and XAS spectral line shape

Figure 2(b) shows a Ni L-edge reflectivity spectrum obtained without laser excitation. The spectral line shape compares well with literature measurements on similar NiO/MgO(001) samples.[7, 28, 31] The dominant feature is a sharp peak at the Ni $L_3$-edge and a double-peak structure at the Ni $L_2$-edge. We note that the width of the sharp $L_3$-edge peak is limited by the energy resolution of the down-stream spectrometer grating. As mentioned above, the energy resolution of the spectrometer grating was 0.33 eV/pixel and the corresponding energy step size can be seen for the flat top of the $L_3$-peak. In the following we focus mainly on the $L_2$-edge since (i) the spectral features are broader and, therefore, contain a sufficient amount of energy steps, and (ii) XMLD measurements have mainly been done using the double-peak structure that gives rise to a characteristic derivative-like XMLD line shape.[7] Since XMLD measurements are typically performed in absorption, we performed ground-state reference XAS measurements monitoring the sample drain current for various grazing incidence angles between 7°-26°, as shown in Fig. 4(a). In all the XAS spectra, one can see the crystal field induced peak splitting at the nickel $L_2$-edge whose energy positions compare well with the shoulder at 873.3 eV and the main peak at 874.5 eV observed in the reflectivity spectra. However, the respective peak intensities differ for XAS and reflectivity. While the XAS spectra correspond to the imaginary part of the X-ray dielectric constant, both imaginary and real parts of the X-ray dielectric constant contribute to the reflectivity. In principle, the real part of the dielectric constant can be computed from the imaginary part using Kramers-Kronig transformation, and consequently, the X-ray reflectivity spectra can be calculated using Fresnel's equation for a specific polarization.[31]

The XAS spectra in Fig. 4(a) show the typical line shape variations that are expected for $L_2$-edge XMLD of NiO/MgO(001).[7] As the incidence angle increases, the intensity of the second peak decreases gradually. The dip between the two peaks slightly shifts towards higher photon energy as the incident angle increases. There is also a red shift of the slope on the right side of the second peak along with increasing incident angles. To calculate the X-ray reflectivity spectrum, we chose the XAS spectrum at a grazing angle of 7° in Fig. 4(a), as it is the closest angle to the one in our time-resolved reflectivity measurement at FLASH. We follow ref. [31] to calculate the reflectivity spectrum. Briefly, (i) use the XAS spectrum as the input to calculate the extinction coefficient $k$, which is the imaginary part of the complex refractive index; (ii) compute the real part $n$ of the complex refractive index from the imaginary part $k$ using Kramers-Kronig transformation; (iii) calculate the reflectivity $R$ using the Fresnel equation for p-polarized light, which is the polarisation in our reflectivity experiment. The Kramers-Kronig transformation is performed using Maclaurin's formula.[32, 33] This results in similar line shapes of the X-ray dielectric constant compared to the reported values in ref. [31]. However, their magnitude was rescaled to the literature values to take the different detection efficiency into account. Figures 4(b, c) show the imaginary and real parts of the X-ray dielectric constant, respectively. From the comparison in Fig. 4(d), one can see reasonable agreement between the measured (blue line) and the calculated (red line) X-ray reflectivity spectra. The lower intensity in the measured reflectivity around 878 eV can occur due to lower flux from the FEL since this energy is close to the end of the scan range.

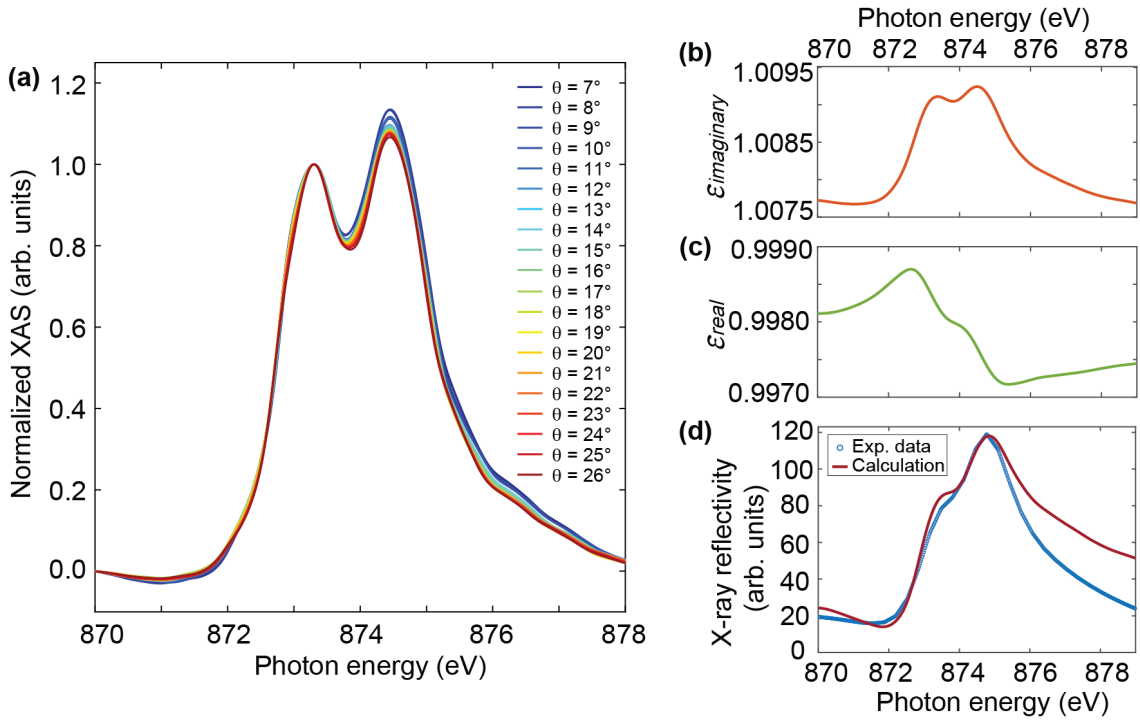

**Fig. 4** (a) Static XAS spectra measured at the nickel $L_2$-edge at different grazing angles. Each XAS spectrum was normalised to the first peak at 873.3 eV. (b) The imaginary part of the dielectric constant, which was calculated using the XAS spectrum at a grazing angle of 7°. (c) The real part of the dielectric constant, which was calculated from the imaginary part in (b) using Kramers-Kronig transformation. (d) Comparison of the experimental X-ray reflectivity spectrum measured at FLASH (blue line) and the calculated reflectivity (red line).

### 3.2    Temporal evolution of AFM order visualized at the nickel $L_2$-edge

The double-peak multiplet structure of the nickel $L_2$-edge has been used extensively to probe the NiO AFM order in thermal equilibrium. The interplay between AFM order, crystalline field effects and spin-orbit coupling gives rise to a characteristic XMLD line shape depending on the alignment of the X-ray polarisation and the AFM moment orientation.[7, 16, 17] In static measurements, the sign of the XMLD line shape can be reversed due to the angle between the X-ray polarisation and the crystal orientation[16] or the subtraction of the X-ray signal between different AFM domains.[17] Here we describe the first NiO XMLD measurement in the time-domain caused by pump-induced changes to the AFM order.

Figure 5(a) shows the static Ni $L_2$-edge X-ray reflectivity of NiO. The difference in reflectivity between laser on and laser off is shown in Fig. 5(b) for two time-delay regions, averaged for the first 400 fs (orange line) and between 400 fs and 2.4 ps (cyan line). The difference reflectivity spectra as a function of time delay and photon energy are shown as a colour map in Fig. 5(c). From Fig. 5(b), one can see that the spectral difference line shape evolves with time delay. After 400 fs the displayed difference spectra in Fig. 5(b) are essentially indistinguishable from what is expected for XMLD in thermal equilibrium.[7, 16, 17] Moreover, the observed positive-negative spectral difference line shape with increasing photon energy demonstrates that a reduction of AFM order takes place.[7] Interestingly, the spectral difference line shape is very different for the first 400 fs. It is characterised by a positive peak at lower photon energy and a negligible negative tail for the high photon-energy multiplet peak. This is an indication of a non-thermal state of the laser-excited spin system, which means the spin system has not reached a local thermal equilibrium. Around 400 fs the difference spectrum changes its line shape which remains unaltered afterwards, as can be clearly seen in the difference map of Fig. 5(c). This implies that the system reaches a local metastable equilibrium after 400 fs. It is conceivable that the XMLD line shape during the first 400 fs is also influenced by electronic excitations that could alter the effective local crystalline field experienced by the nickel ions. Crystalline field changes in monolayer thin NiO films have been reported to lead to changes in the $L_2$ multiplet peaks different from XMLD.[34]

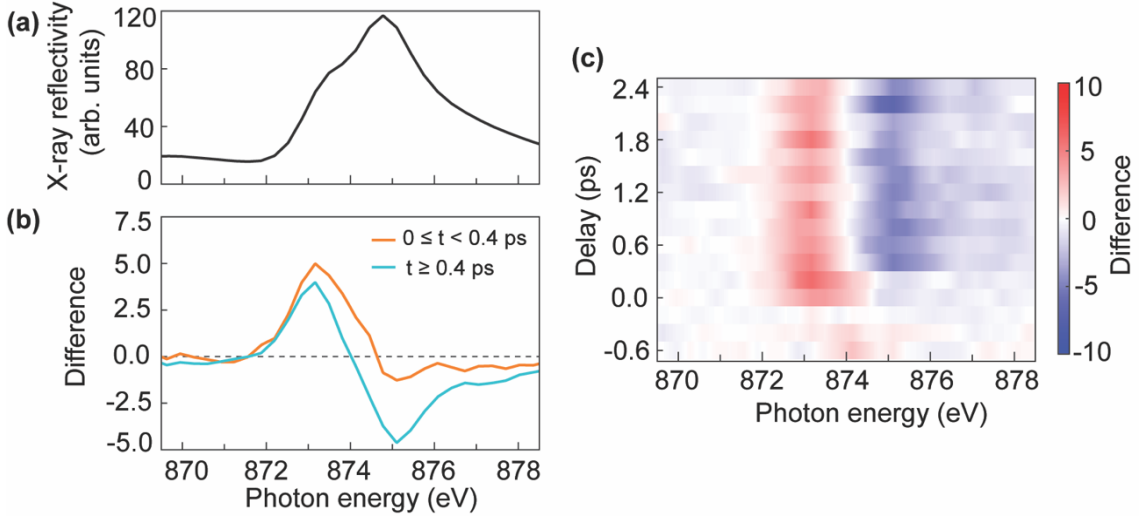

**Fig. 5** (a) The static X-ray reflectivity spectrum at the nickel $L_2$-edge. (b) The difference spectra of the X-ray reflectivity with laser and without laser at transient time (0 ≤ t < 400 fs, orange line) and at longer delays (t ≥ 400 fs, cyan line). (c) The map of the difference spectra as a function of delay and photon energy.

Since the spin system reaches a local equilibrium after 400 fs, it is of interest to evaluate the temperature change of the spin system. It has been observed that the temperature dependence of the ratio between the two multiplets peaks at the nickel $L_2$-edge follows a Brillouin-Langevin function.[7] We observe a 7% change in the cyan XMLD spectrum of Fig. 5(b). This implies that we can use a linear approximation to evaluate the spin-temperature-change of the system. We convert the $L_2$ multiplet peak ratio measured in reflectivity to absorption as outlined in section 3.1 and use the measured temperature dependence in ref. [7] to obtain an effective spin temperature increase of 65 ± 5 K, relative to the room temperature starting condition before optical excitation. The uncertainty is estimated from the photon statistics.

This observation of a significant reduction of the AFM order in NiO thin films is surprising especially since a negligible change was reported in a bulk sensitive magnetic scattering experiment performed under very similar excitation conditions.[6] In the following we discuss if optical excitation can indeed deposit enough energy to drive a comparable temperature increase. We evaluated the absorption coefficient of our NiO film at 800 nm to be around 1860 cm$^{-1}$. The refractive index of NiO at the excitation wavelength is 2.35 and the reflection of the sample is about 20% for p-polarized light.[35] Using the Beer-Lambert law and the magnetic specific heat of NiO,[36] we estimated the maximum temperature rise of the spin system to be ~150 K. This represents an upper limit since the magnetic specific heat increases rapidly as the temperature approaches $T_N$.[36] We alternatively estimated the lattice temperature rise base on a previous ultrafast electron diffraction measurement.[3] Assuming that at the slightly higher pump photon energy used in ref. [3] the optical absorption is of at least similar strength, we can extrapolate the measured lattice temperature in ref. [3] to ~85 K given our pump fluence. We note that in ref. [3] no optical absorption data for the measured sample was available and the available literature data provided an insufficient heat deposition leading the authors in ref. [3] to conclude that two-photon above-gap excitation was the actual driving force. However, our lower photon energy would require three-photon events for above-gap absorption. We can conclude that the enhanced optical absorption in thin NiO films is sufficient to explain our observed level of AFM order reduction.

### 3.3 Band gap renormalisation visualized at the oxygen K-edge

Figure 6 displays the results of transient reflectivity spectra measured at the oxygen K-edge. We focus on the lowest photon energy feature as this is the one that defined the upper band gap. However, we corroborated in static measurements that our samples display the typical three-peak structure between 525 eV and 545 eV expected for NiO.[37-39] Figure 6(a) shows the static X-ray reflectivity as a reference for the map of the difference spectra between laser on and laser off in Fig. 6(b). The pre-edge dip at 531.5 eV in Fig. 6(a) is related to the reflection geometry, which can also be seen from the calculated spectrum in Fig. 4(d). The evolution of the difference spectra shown in Fig. 6(b) is plotted using a temporal binning of 200 fs to improve the signal-to-noise ratio (SNR). The positive signal region centred around 532 eV in Fig. 6(b) is an indication of a red-shift ($\approx$ 60 meV) of the oxygen K-edge. The amount of the red-shift will be discussed in detail in the next section in combination with first-principles modelling.

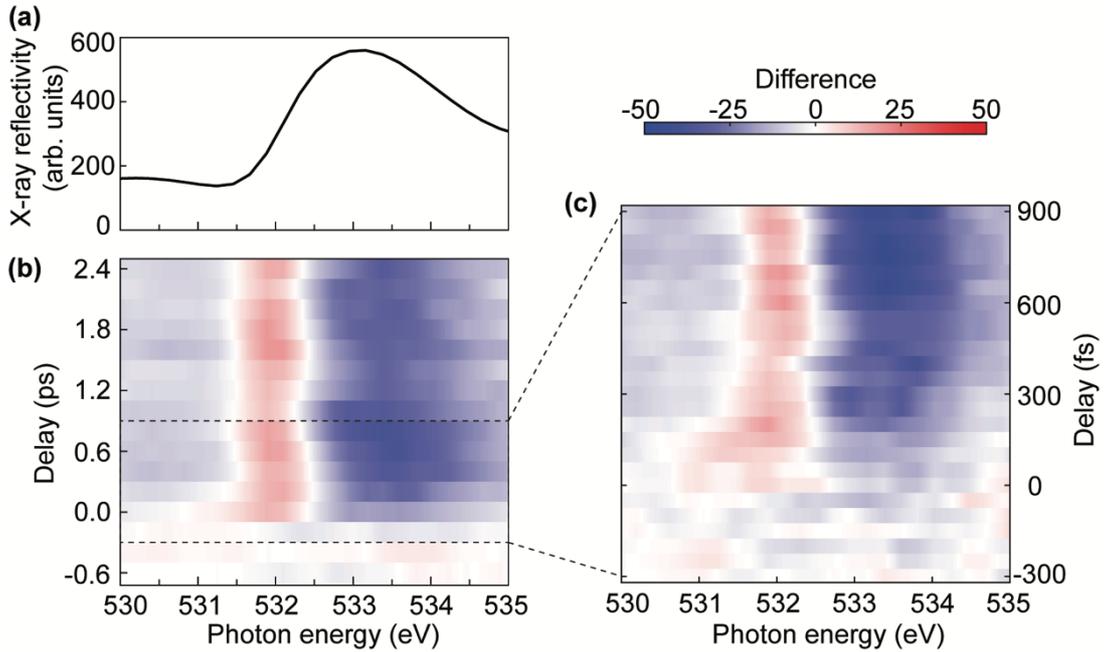

**Fig. 6** (a) The static X-ray reflectivity spectrum at the oxygen K-edge. (b) The map of the difference spectra of the X-ray reflectivity with laser and without laser as a function of delay and photon energy. (c) The map of laser-induced difference in X-ray reflectivity as a function of delay and photon energy at a fixed undulator gap. The delay step size is 50 fs.

The most prominent feature visible in the difference map of Fig. 6(b) is clearly the red-shifted absorption edge, indicated by the red intensity increase near 532 eV photon energy and a broader blue decrease above the edge (533-534 eV). A gap reduction following sub-gap optical excitation has been observed in the charge-transfer insulator, $La_2CuO_4$,[40] and it is tempting to relate our observation to the same effect. However, strictly speaking we observe in the present experiment only the red-shifted upper gap-edge while a lower gap-edge shift would have to be assess e.g. in complementary time-resolve photoemission experiments. The gap renormalisation in ref. [40] was assigned to the coupling of optical excitation to a bosonic field, although it remained unclear if this is composed of phonons or magnons. A possible magnetic origin of the red-shifted NiO upper gap-edge becomes apparent when we analyse the observed oscillations visible in Fig. 6(b).

The oscillation of the positive difference signal centred around 532 eV is compatible with a period of ~1 ps, which corresponds to a frequency of ~1 THz. This frequency is close to the eigenfrequency of the out-of-plane magnon mode, which has been extensively studied in both optical and THz spectroscopy measurements.[14, 19, 41-49] There are two magnon eigenmodes: in-plane and out-of-plane modes. For the in-plane mode, the AFM vector between two adjacent ferromagnetic planes is modulated along the [1-10] direction, which lies in the (111) plane. For the out-of-plane mode, the modulation of the AFM vector is along the [111] direction, which is perpendicular to the ferromagnetic plane.[20] Due to the large magnetic anisotropy along the easy axis (i.e. [111] direction), the out-of-plane mode has a much higher eigenfrequency than the in-plane mode. Both of these two magnon modes are excitable by ultrashort optical laser pulses and have been detected at around 0.14 THz[20, 46] and 1.07 THz[20, 42, 45-47, 49] for the in-plane mode and the out-of-plane mode, respectively. The mode also displays a temperature dependence with a frequency reduction as the Néel temperature is approached.[14, 42, 49] We note that also the observed phase of the oscillation in Fig. 6(b) matches that reported for the out-of-plane magnon mode.[20]

Figure 6(c) shows the oxygen K-edge difference reflectivity signal up to 900 fs with 50 fs step size. The measurements were performed at a fixed undulator gap corresponding to a nominal photon energy of 533 eV. The displayed photon energy region therefore represents the full bandwidth of the undulator emission. The line shape of the X-ray reflectivity (not shown) is similar to the one displayed in Fig. 6(a), however, the best SNR is obtained for the undulator maximum (533 eV). Figure 6(c) shows, with a much better time resolution than Fig. 6(b), that the red-shift of the upper gap-edge is preceded by the appearance of transient states much further into the band-gap. The temporal duration of such transient mid-gap states seems to essentially be determined by our optical pulse duration convoluted with the temporal resolution. This reflects to some degree earlier measurements at pump photon energies off-resonant to any d-d transitions in the NiO band-gap.[9] It is conceivable that the appearance of transient mid-gap states are a consequence of the strong optical driving field. However, with off-resonance excitation no significant lasting band-gap renormalisation was observed.[9] The red-shift of the upper gap-edge is therefore likely a consequence of the energetic proximity of our excitation frequency to d-d transitions that may facilitate the coupling to spin excitations.[50]

### 3.4 A simple model for the band gap renormalisation in NiO

With the evidence of the laser-induced XMLD line shape changes at the nickel $L_2$-edge (section 3.2) and the observation of the 1 THz magnon oscillation in the NiO upper gap-edge red-shift (section 3.3), it is of interest to investigate how changes in the spin orientation influence the band gap in NiO. Therefore, we performed first-principles calculations of the NiO band structure with different canting angles of the $Ni^{2+}$ spins, to mimic various spin temperatures. The super-exchange and correlation in the calculations were taken into account using local density approximation together with an on-site Hubbard U term in the Hamiltonian. Although dynamical correlations have been pointed out to be important for NiO, we adopt for simplicity here the LDA+U formalism, that is based on single determinant states. A detailed description of the theoretical modelling can be found in ref. [9]. The canting angles at different temperatures were derived from the reduction of the magnetic moment of the $Ni^{2+}$ spins.[15]

Figure 7(a) shows the band structure calculated with $Ni^{2+}$ spins oriented along the [11-2] direction. The conduction band is composed of essentially dispersionless bands of Ni 3d character and dispersive bands consisting of hybridised O 2p and Ni 4s states. The upper gap-edge is composed of the latter and is marked by the light blue shaded rectangular in Fig. 7(a). The green and orange dots in Fig. 7(a) represent the experimental $q$ values probed in our time-resolved X-ray reflectivity measurements. Figure 7(b) shows the influence of canted magnetic moments on adjacent Ni atoms along the [111] on the dispersive band near the upper gap-edge. The chosen spin canting pattern is that of the 1 THz magnon mode [20] (see Fig. 7(c)). Initially degenerate bands are split in energy once the spins are canted. This splitting increases as the canting angle becomes larger. The band shifting downwards in energy at the Γ point [000] of the Brillouin zone reduces the NiO band gap as probed by the oxygen K-edge. Although the band structure in Fig. 7 was calculated for frozen magnetic moments, it is straightforward to see that for oscillating spins the upper gap-edge position would also be modulated by the magnon frequency.

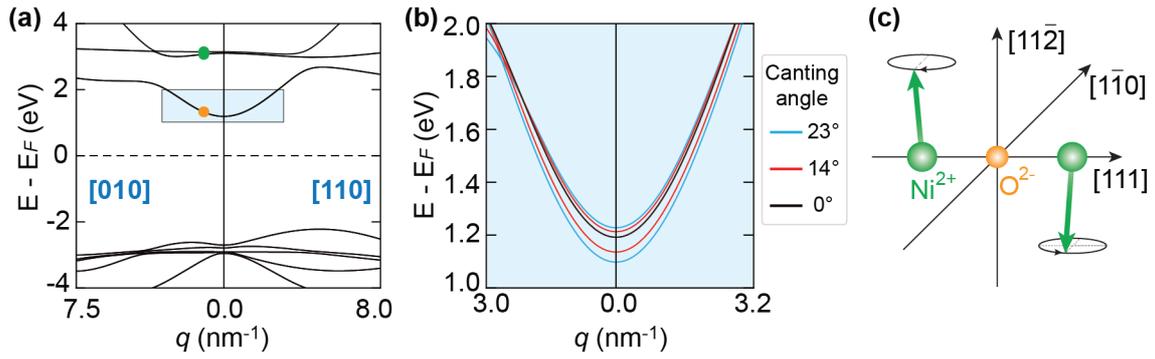

**Fig. 7** (a) First-principles calculations of the NiO band structure along two high symmetry lines:[9] [010] and [110]. The green and orange dots along the [010] direction denote the experimental wavevectors that were probed in the time-resolved X-ray reflectivity measurement. (b) Zoom-in view of the band structure marked within the light blue shaded rectangular in (a). The band structure was calculated for different canting angles of the $Ni^{2+}$ spins. (c) Illustration of the canting angle of the $Ni^{2+}$ spins in NiO unit cell that was employed in the calculations. The canting angle is defined as the angle between the $Ni^{2+}$ spins and the plane that is spanned by [11-2] and [111] directions. The spin canting pattern is the same as the out-of-plane mode in ref. [20]. The $Ni^{2+}$ and the $O^{2-}$ ions are shown with green and orange spheres, respectively.

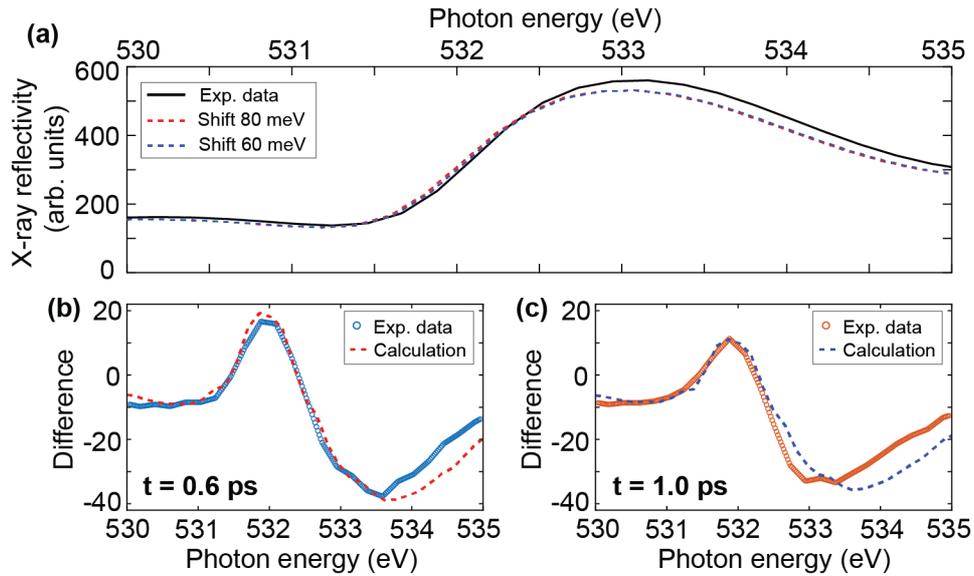

**Fig. 8** (a) The black line represents the static X-ray reflectivity spectrum acquired at FLASH. The dashed red and blue lines represent the shifted spectra by 80 meV and 60 meV, respectively. (b) Comparison of the experimental difference spectrum at 0.6 ps after laser excitation and the calculated spectrum that is obtained by subtracting the black line from the dashed red line in (a). (c) Comparison between the experimental data at 1.0 ps after laser excitation and calculated difference spectrum that is the subtraction of the black line and the dashed blue line in (a).

In the following we quantify the redshift of the upper gap-edge and separate the coherent oscillatory part from an incoherent redshift. Figure 8(a) compares the measured ground-state X-ray reflectivity spectrum (solid line) to the redshifted experimental spectra for 60 (blue dashed line) and 80 meV (red dashed line) shifts. In addition to the redshift, we also observe experimentally a 5% reduction to the spectral peak intensity likely related to the blueshift of some initially degenerate bands as seen in Fig. 7(b). This results in a good agreement with experiment for a redshift of 80 ± 10 meV at 0.6 ps time delay (see Fig. 8(b)) and 60 ± 10 meV at 1.0 ps time delay (see Figs. 8(c)). We note that for increasing photon energy the agreement becomes less good possibly due to the influence of additional O K-edge spectral peaks neglected here. From Fig. 6(b) one can see that the positive signal around 532 eV is composed of an overall intensity increase and an oscillatory intensity variation. The overall intensity increase corresponds to an incoherent redshift of 70 meV. The amplitude of the coherent oscillatory redshift related to the 1 THz magnon mode is approximately 10 meV. Using the calculations in Fig. 7(b) we can estimate the corresponding magnon precession angle to ~2.5°. We note that the reduction of AFM due to such large spin precession amplitude would amount to about 28% of the spin temperature increase we observed in section 3.2.

## 4    Conclusions

In conclusion, we performed time-resolved resonant X-ray reflectivity measurements in NiO following femtosecond 1.5 eV sub-gap excitation. The temporal evolution of the X-ray reflectivity spectra was monitored at the Ni $L_2$-edge and the O K-edge to probe the dynamics of the AFM order and the band gap renormalisation, respectively.

At the Ni $L_2$-edge, we observed laser-induced XMLD spectral line shape for time delays longer than ~400 fs after laser excitation. By comparing the transient XMLD spectra with temperature-dependent ground-state measurements, we extracted a spin temperature rise of 65 ± 5 K. The transient XMLD line shape demonstrates a significant reduction of the AFM order in NiO. The change of the spin temperature is consistent with what is expected from optical absorption measurements. At the O K-edge, we observe the red-shift of the upper band gap edge across the measured time delay range up to 2.4 ps. This upper gap-edge renormalisation is accompanied by a coherent oscillation with ~1 THz frequency. During the optical driving field we find evidence for the formation of transient mid-gap states that were also observed previously for optical frequencies off-resonant to d-d excitations.[9]

We finally discuss a ground-state model of coupling Ni spin precession to the electronic band-structure. Using first-principles calculations, we find that the O 2p derived states forming the upper gap-edge shift in energy as the Ni magnetic moments precess as given by the 1 THz magnon mode. This even allows us to model the observed O K-edge measurements.

The high-quality X-ray spectroscopy measurements presented here have become possible due to the implementation of a multiple X-ray beams experimental setup that enables us to all but eliminate the intensity and photon-energy fluctuations inherent to SASE FELs. This opens up new possibilities to investigate the ultrafast dynamics of electronic structure and the magnetic ordering phenomena across a wide range of materials.

## Author contributions

M. B. and H. A. D. designed the experiments. X. W., R. Y. E., I. V., D. T., J. O. S., S. D., G. B., R.-P. W., M. K., F. K., W. B., M. B. and H. A. D. performed the time-resolved X-ray reflectivity experiment and on-site data analysis. R. K. and I. V. performed the optical spectroscopy experiment. C. S.-L. performed the static XAS experiment. X. W. conducted the post data analysis. O. G. conducted the theoretical calculations. X. W., R. Y. E., I. V., D. T., V. S., A. Y., O. G., F. P., O. E., M. B., and H. A. D. discussed the results. X. W. wrote the first draft of the manuscript. X. W. and H. A. D. revised the manuscript with inputs from all other authors.

## Conflicts of interest

There are no conflicts to declare.

## Acknowledgements

The authors acknowledge FLASH in Hamburg, Germany, for provision of X-ray free-electron laser beamtime at beamline FL24 and thank the machine group and facility staff for their assistance during the beamtime. The beamtime was allocated for proposal F-20191553 EC. The research work at FLASH has been supported by the project CALIPSOplus under the Grant Agreement 730872 from the EU Framework Programme for Research and Innovation HORIZON 2020. X. W., D. T., V. S., and H. A. D. acknowledge support from the Swedish Research Council (VR), Grants 2017-06711 and 2018-04918. A. Y. acknowledges support from the Carl Trygger Foundation. O. G. acknowledges support from the Strategic Research Council (SSF) grant ICA16-0037 and the Swedish Research Council (VR) grant 2019-03901. The computations were enabled by resources provided by the Swedish National Infrastructure for Computing (SNIC), partially funded by the Swedish Research Council through grant agreement no. 2018-05973. R. Y. E., J. O. S., and M. B. were supported by the Helmholtz Association through grant VH-NG-1105. I. V. acknowledges support from Slovene Ministry of Science (Raziskovalci-2.1-IJS-952005). R.-P. W. acknowledges support by German Ministry of Education and Research, grant number 05K19GU2. We thank Prof. Arne Roos for his help in the optical spectroscopy measurement.

## References

1. T. Jungwirth, X. Marti, P. Wadley and J. Wunderlich, *Nature Nanotechnology*, 2016, **11**, 231-241.
2. V. Baltz, A. Manchon, M. Tsoi, T. Moriyama, T. Ono and Y. Tserkovnyak, *Reviews of Modern Physics*, 2018, **90**, 015005.
3. Y. W. Windsor, D. Zahn, R. Kamrla, J. Feldl, H. Seiler, C.-T. Chiang, M. Ramsteiner, W. Widdra, R. Ernstorfer and L. Rettig, *Physical Review Letters*, 2021, **126**, 147202.
4. M. A. Peck and M. A. Langell, *Chemistry of Materials*, 2012, **24**, 4483-4490.
5. A. M. Balagurov, I. A. Bobrikov, S. V. Sumnikov, V. Y. Yushankhai, J. Grabis, A. Kuzmin, N. Mironova-Ulmane and I. Sildos, *Physica Status Solidi (B)*, 2016, **253**, 1529-1536.
6. L. Huber, A. Ferrer, T. Kubacka, T. Huber, C. Dornes, T. Sato, K. Ogawa, K. Tono, T. Katayama, Y. Inubushi, M. Yabashi, Y. Tanaka, P. Beaud, M. Fiebig, V. Scagnoli, U. Staub and S. L. Johnson, *Physical Review B*, 2015, **92**, 094304.
7. D. Alders, L. H. Tjeng, F. C. Voogt, T. Hibma, G. A. Sawatzky, C. T. Chen, J. Vogel, M. Sacchi and S. Iacobucci, *Physical Review B*, 1998, **57**, 11623.
8. L.-C. Duda, T. Schmitt, M. Magnuson, J. Forsberg, A. Olsson, J. Nordgren, K. Okada and A. Kotani, *Physical Review Letters*, 2006, **96**, 067402.
9. O. Grånäs, I. Vaskivskyi, P. Thunström, S. Ghimire, R. Knut, J. Söderström, L. Kjellsson, D. Turenne, R. Y. Engel, M. Beye, J. Lu, A. H. Reid, W. Schlotter, G. Coslovich, M. Hoffmann, G. Kolesov, C. Schüßler-Langeheine, A. Styervoyedov, N. Tancogne-Dejean, M. A. Sentef, D. A. Reis, A. Rubio, S. S. P. Parkin, O. Karis, J. Nordgren, J.-E. Rubensson, O. Eriksson and H. A. Dürr, 2020, arXiv:2008.11115.
10. D. Betto, Y. Y. Peng, S. B. Porter, G. Berti, A. Calloni, G. Ghiringhelli and N. B. Brookes, *Physical Review B*, 2017, **96**, 020409(R).
11. G. Lefkidis and W. Hübner, *Physical Review Letters*, 2005, **95**, 077401.
12. A. K. Cheetham and D. A. O. Hope, *Physical Review B*, 1983, **27**, 6964.
13. Y. Gao, Q. Sun, J. M. Yu, M. Motta, J. McClain, A. F. White, A. J. Minnich and G. K.-L. Chan, *Physical Review B*, 2020, **101**, 165138.
14. J. Nishitani, K. Kozuki, T. Nagashima and M. Hangyo, *Applied Physics Letters*, 2010, **96**, 221906.
15. N. Rinaldi-Montes, P. Gorria, D. Martínez-Blanco, A. B. Fuertes, I. Puente-Orench, L. Olivi and J. A. Blanco, *AIP Advances*, 2016, **6**, 056104.
16. E. Arenholz, G. van der Laan, R. V. Chopdekar and Y. Suzuki, *Physical Review Letters*, 2007, **98**, 197201.
17. H. Ohldag, G. van der Laan and E. Arenholz, *Physical Review B*, 2009, **79**, 052403.
18. J. Xu, C. Zhou, M. Jia, D. Shi, C. Liu, H. Chen, G. Chen, G. Zhang, Y. Liang, J. Li, W. Zhang and Y. Wu, *Physical Review B*, 2019, **100**, 134413.
19. T. Satoh, S.-J. Cho, R. Iida, T. Shimura, K. Kuroda, H. Ueda, Y. Ueda, B. A. Ivanov, F. Nori and M. Fiebig, *Physical Review Letters*, 2010, **105**, 077402.
20. C. Tzschaschel, K. Otani, R. Iida, T. Shimura, H. Ueda, S. Günther, M. Fiebig and T. Satoh, *Physical Review B*, 2017, **95**, 174407.
21. T. Kachel, F. Eggenstein and R. Follath, *Journal of Synchrotron Radiation*, 2015, **22**, 1301-1305.
22. R. Newman and R. M. Chrenko, *Physical Review*, 1959, **114**, 1507.
23. I. G. Austin, B. D. Clay and C. E. Turner, *Journal of Physics C: Solid State Physics*, 1968, **1**, 1418.
24. G. R. Rossman, R. D. Shannon and R. K. Waring, *Journal of Solid State Chemistry*, 1981, **39**, 277-287.
25. T. Tsuboi and W. Kleemann, *Journal of Physics: Condensed Matter*, 1994, **6**, 8625.

26. V. I. Sokolov, V. A. Pustovarov, V. N. Churmanov, V. Y. Ivanov, N. B. Gruzdev, P. S. Sokolov, A. N. Baranov and A. S. Moskvin, *Physical Review B*, 2012, **86**, 115128.
27. M. Beye, R. Y. Engel, J. O. Schunck, S. Dziarzhytski, G. Brenner and P. S. Miedema, *Journal of Physics: Condensed Matter*, 2018, **31**, 014003.
28. R. Y. Engel, P. S. Miedema, D. Turenne, I. Vaskivskyi, G. Brenner, S. Dziarzhytski, M. Kuhlmann, J. O. Schunck, F. Döring, A. Styervoyedov, S. S. P. Parkin, C. David, C. Schüßler-Langeheine, H. A. Dürr and M. Beye, *Applied Sciences*, 2020, **10**, 6947.
29. B. L. Henke, E. M. Gullikson and J. C. Davis, *Atomic data and nuclear data tables*, 1993, **54**, 181-342.
30. C. Gahl, A. Azima, M. Beye, M. Deppe, K. Döbrich, U. Hasslinger, F. Hennies, A. Melnikov, M. Nagasono, A. Pietzsch, M. Wolf, W. Wurth and A. Föhlisch, *Nature Photonics*, 2008, **2**, 165-169.
31. D. Alders, T. Hibma, G. A. Sawatzky, K. C. Cheung, G. E. van Dorssen, M. D. Roper, H. A. Padmore, G. van der Laan, J. Vogel and M. Sacchi, *Journal of Applied Physics*, 1997, **82**, 3120-3124.
32. K. Ohta and H. Ishida, *Applied Spectroscopy*, 1988, **42**, 952-957.
33. T. G. Mayerhöfer and J. Popp, *Spectrochimica Acta Part A: Molecular and Biomolecular Spectroscopy*, 2019, **213**, 391-396.
34. M. W. Haverkort, S. I. Csiszar, Z. Hu, S. Altieri, A. Tanaka, H. H. Hsieh, H.-J. Lin, C. T. Chen, T. Hibma and L. H. Tjeng, *Physical Review B*, 2004, **69**, 020408(R).
35. M. M. El-Nahass, M. Emam-Ismail and M. El-Hagary, *Journal of Alloys and Compounds*, 2015, **646**, 937-945.
36. R. J. Radwanski and Z. Ropka, *Acta Physica Polonica-Series A General Physics*, 2008, **114**, 213-218.
37. H. Y. Peng, Y. F. Li, W. N. Lin, Y. Z. Wang, X. Y. Gao and T. Wu, *Scientific Reports*, 2012, **2**, 442.
38. R. J. O. Mossanek, G. Domínguez-Cañizares, A. Gutiérrez, M. Abbate, D. Díaz-Fernández and L. Soriano, *Journal of Physics: Condensed Matter*, 2013, **25**, 495506.
39. J. Y. Zhang, W. W. Li, R. L. Z. Hoye, J. L. MacManus-Driscoll, M. Budde, O. Bierwagen, L. Wang, Y. Du, M. J. Wahila, L. F. J. Piper, T.-L. Lee, H. J. Edwards, V. R. Dhanak and K. H. L. Zhang, *Journal of Materials Chemistry C*, 2018, **6**, 2275-2282.
40. F. Novelli, G. De Filippis, V. Cataudella, M. Esposito, I. Vergara, F. Cilento, E. Sindici, A. Amaricci, C. Giannetti, D. Prabhakaran, S. Wall, A. Perucchi, S. Dal Conte, G. Cerullo, M. Capone, A. Mishchenko, M. Grüninger, N. Nagaosa, F. Parmigiani and D. Fausti, *Nature Communications*, 2014, **5**, 5112.
41. S. Kovalev, Z. Wang, J.-C. Deinert, N. Awari, M. Chen, B. Green, S. Germanskiy, T. V. A. G. de Oliveira, J. S. Lee, A. Deac, D. Turchinovich, N. Stojanovic, S. Eisebitt, I. Radu, S. Bonetti, T. Kampfrath and M. Gensch, *Journal of Physics D: Applied Physics*, 2018, **51**, 114007.
42. T. Moriyama, K. Hayashi, K. Yamada, M. Shima, Y. Ohya and T. Ono, *Physical Review Materials*, 2020, **4**, 074402.
43. T. Kampfrath, A. Sell, G. Klatt, A. Pashkin, S. Mährlein, T. Dekorsy, M. Wolf, M. Fiebig, A. Leitenstorfer and R. Huber, *Nature Photonics*, 2011, **5**, 31-34.
44. J. Nishitani, T. Nagashima and M. Hangyo, *Physical Review B*, 2012, **85**, 174439.
45. S. Baierl, J. H. Mentink, M. Hohenleutner, L. Braun, T.-M. Do, C. Lange, A. Sell, M. Fiebig, G. Woltersdorf, T. Kampfrath and R. Huber, *Physical Review Letters*, 2016, **117**, 197201.
46. T. Kohmoto, T. Moriyasu, S. Wakabayashi, H. Jinn, M. Takahara and K. Kakita, *Journal of Infrared, Millimeter, and Terahertz Waves*, 2018, **39**, 77-92.
47. Z. Wang, S. Kovalev, N. Awari, M. Chen, S. Germanskiy, B. Green, J.-C. Deinert, T. Kampfrath, J. Milano and M. Gensch, *Applied Physics Letters*, 2018, **112**, 252404.
48. O. V. Chefonov, A. V. Ovchinnikov, C. P. Hauri and M. B. Agranat, *Optics Express*, 2019, **27**, 27273-27281.
49. T. Moriyama, K. Hayashi, K. Yamada, M. Shima, Y. Ohya, Y. Tserkovnyak and T. Ono, *Physical Review B*, 2020, **101**, 060402(R).
50. R. Gómez-Abal, O. Ney, K. Satitkovitchai and W. Hübner, *Physical Review Letters*, 2004, **92**, 227402.